\documentclass{jetp}
\usepackage{amssymb}

\twocolumn

\begin{document}

\English

\title{Order parameter in electron systems: its fluctuations and oscillations}
\author{Konstantin B.}{Efetov}
\email{Konstantin.B.Efetov@ruhr-uni-bochum.de}

\affiliation{Ruhr University Bochum, Faculty of Physics and Astronomy, Bochum, 44780, Germany}
\affiliation{National University of Science and Technology \textquotedblleft MISiS\textquotedblright, Moscow, 119049, Russia}
\affiliation{International Institute of Physics, UFRN, 59078-400 Natal, Brazil}
\date{today}

\abstract{The concept of the order parameter is extremely useful in physics.
Here, I discuss extensions of this concept to cases when the order parameter
is no longer a constant but fluctuates or oscillates in space and time. This
allows one to describe in an unified manner diverse physical phenomena
including coexisting superconductivity and insulators in
(quasi)one-dimensional systems, superconductivity and Coulomb blockade in
granular superconductors and Josephson networks, Anderson localization and
mesoscopic effects in disordered and chaotic systems, and thermodynamic
quantum time-space crystals.}

\maketitle

\section{Introduction}

The concept of the order parameter introduced by Landau in his theory of
phase transitions \cite{landau} plays the central role in condensed matter
and statistical physics. It has become clear later with development of the
scaling theory that, in order to describe a phase transition, one should
integrate over the order parameter with a weight determined by a
Ginzburg-Landau-Wilson free energy functional \cite{wilson}. This field of
research has attracted a lot of interest at the Landau Institute in
particular because scaling ideas had been proposed previously by
Patashinskii and Pokrovskii \cite{pp} and by Kadanoff \cite{kadanoff}, and
first renormalization group study of the phase transitions had been
performed by Larkin and Khmel'nitski \cite{lk}. I started my scientific
activity at the Landau Institute in the time when all these ideas had just
appeared and one could hear hot discussions on that topic.

Simultaneously, it was getting clear that fluctuations of the order
parameter could be very important not only near a phase transition by also
in low-dimensional systems. The dimension of the system was determined by
geometry of the sample, while the dimension of the electron bands was not
always important. My attention to this class of problems was drawn by Tolya
Larkin, with whom I made my PhD. We have understood quite generally that the
behavior of two-point correlation functions of one-dimensional electron
systems at large distances or times were completely determined by sound-like
gapless quantum fluctuations \cite{efetov1975}. Although many
thermodynamical physical quantities could be calculated microscopically
using rather sophisticated Bethe Ansatz methods, one could not determine the
correlation functions using that technique. Instead, we have demonstrated
that it was sufficient to compute correlations of superconducting or
insulating order parameters taking into account their fluctuations.
Actually, this was one of the first step in the subsequent development of
powerful bosonization techniques \cite{haldane} (for review, see, e.g. Ref.
\cite{tsvelik}).

The idea that many interesting effects can be efficiently described by
considering low energy fluctuations of the order parameter motivated me
later to study physics of granular superconductors \cite{efetov1980}. In
these materials, the superconductivity in a single grain can be well
described by a phase of the order parameter fluctuating in time. The modulus
of the order parameter was assumed to be a constant and there was no need to
consider space variations of the phase inside the grain. In order to
describe the macroscopic superconductivity in the array of the grains one
had to account for the Josephson coupling between the grains. A Coulomb
interaction turns out to be very important enhancing the phase fluctuations
in time and eventually leading to a superconductor-insulator transition
dubbed later `Coulomb blockade'. Since then, I have made several other works
on granular superconductors, and one can get more details in the review \cite%
{beloborodov}. Physics of granular superconductors is the same as physics of
artificially designed Josephson networks and the description developed in
Ref. \cite{efetov1980} has been used and further developed later in a huge
number of publications.

Surprisingly, the idea of fluctuations of the order parameter turned out to
be fruitful in problems of Anderson localization and mesoscopics.
Originally, it was suggested in a `bosonic' replica reformulation of models
with disorder \cite{wegner} followed by `fermionic' replica representation
\cite{elk}. Within this approach one reduces summation of certain classes of
diagrams (so called `diffusons' and `cooperons') to study of fluctuations of
a matrix that looked like an order parameter. The matrix looked formally as
an order parameter and a `free energy functional' looked very similar to the
one describing fluctuations of the phase in superconductor. The free energy
functional had a form of a non-linear $\sigma $-model.

It turned out very soon that, although the $\sigma $-model allowed one to
perform very efficiently perturbation theory and renormaliztion group
calculations, it was not possible to do non-perturbative calculations. In
order to circumvent this difficulty I have derived a supermatrix $\sigma $%
-model \cite{efetov1982} that was free of these problems and allowed one to
perform essentially non-perturbative calculations. I continued to work in
this direction for quite a long time because the method worked not only for
the localization problems but also in mesoscopics, quantum chaos, random
matrices, etc. \cite{efetov1983,book}.

The supermatrix `order parameter' $Q$ that appears in the $\sigma $-model
approach must be averaged with the free energy functional, and it has no
physical sense without carrying out this procedure. Actually, the average $%
\left\langle Q\right\rangle $ with the action of the $\sigma $-model is not
an interesting quantity because it is the average density of state and the
latter quantity is a smooth function of energy. Conductivity, level
correlations, density-density correlations, etc. can be written in terms of
a product of several $Q$ like, e.g. $\left\langle QQ\right\rangle .$
Therefore, the matrix $Q$ is not an order parameter in its usual sense. In
order to obtain an interesting physical quantity, one should integrate a
product of several $Q$ over all configurations.

To my great surprise, I have encountered a rather similar situation in my
investigation of a possibility of existence of a thermodynamically stable
`Time-crystal'. The time crystal is expected to demonstrate an oscillating
behavior of physical quantities in time. The concept of a time-crystal has
been proposed several years ago \cite{wilczek} in a simple model but later
it turned out that the time crystal state proposed there was not the ground
state and therefore could not be stable. Moreover, it was even argued that
the thermodynamically stable quantum crystal could not exist at all \cite%
{watanabe}. The `no-go' theorem has been proven for systems with
time-independent wave functions of the ground state, which is usually the
case.

However, in a recent preprint \cite{efetov2019}, I have suggested and
investegated a model that can undergo a transition into a state with an
order parameter $b$ depending on both real and imaginary times. As a result,
wave functions also depend on the real and imaginary time. The position of
this non-trivial order parameter in time is arbitrary and the averaging over
the positions gives zero. At the same time, the average of a product of the
order parameters can be finite and can be measured experimentally. For
example, the average $\left\langle bb\right\rangle $ can be measured in
quantum scattering experiments.

A more detailed presentation of the results listed in the introduction will
be given in the next sections. It is not a review of publications in several
different fields of physics. I simply want to emphasize considering several
examples that many new results can be obtained using the generalized concept
of the order parameter. Using this approach one can considerably simplify
calculations because it is sufficient to consider large distances without
going into details of band structures and interactions. Very often this
route gives a possibility to solve problems that have not been solved before
and predict new physical phenomena. I have realized the efficiency of this
approach during my years at the Landau Institute and used it later in many
works.

Although I have spent at the Landau Institute as Master and PhD student and
later as researcher only 17 years of my already rather long scientific
carrier, these years were decisive in forming my scientific profile and
scientific tastes. It was typical for scientists working there to develop
general fruitful concepts, and scientific criteria were very high. I myself
have been following these principles all over the years after leaving the
Institute.

Isaak Markovich Khalatnikov played an absolutely outstanding role in
creating the Institute, selecting researchers from different fields. Of
course, there have been other people who played a very important role in the
development of the Institute but his role was unique. The main criterion of
selecting new members of the Institute was their ability to do outstanding
original reasearch, and different political reasons did not play a
considerable role. It is amazing how Isaak Markovich tried to understand
results of all works done at the Institute. Even now he really listens talks
at seminars and meetings of the Scientific Council. He was always extremely
proud of good works performed at the Institute, although it usually did not
mean that he was coauthor of those publications. I definitely have nostalgic
reminiscences about the time of my work at the Institute.

The present paper is organized as follows:

Sections \ref{sec:bosonization}-\ref{sec:supersym} are devoted to a short
review of several works that I have done during my work at the Landau
Institute that demonstrate how one can use the concept of fluctuating order
parameter. Section \ref{sec:bosonization} contains discussion of how one can
calculate non-trivial correlation functions using a hypothesis that their
form is determined by gapless excitations, in Section \ref{sec:granular} it
is shown that Coulomb interaction can destroy the macroscopic
superconductivity and make the system insulating, while Section \ref%
{sec:supersym} is devoted to development of the supersymmetry method for
studying Anderson localization, physics of mesoscopic systems, quantum
chaotic motion, etc. In Section \ref{sec:time} I present results of a recent
work where the thermodynamic quantum time-space crystal is proposed. The
order parameter of a such a crystal oscillates not only in space but also in
both real and imaginary time. Conclusions are made in Section \ref%
{sec:conclusion}.

\section{\label{sec:bosonization}Bosonization of fermionic models in 1D.}

The idea of my first PhD works \cite{eflar1974,efetov1975} with Tolya Larkin
of reducing calculations for rather complicated 1D models of fermions to
simplified models describing low energy fluctuations of the order parameter
was motivated by interest to studying quasi-one-dimensional materials. In
the limit of weak coupling between the chains fluctuations become very
strong and one cannot use conventional mean field theory. In the first work
\cite{eflar1974}, we suggested an idea to study the system by making a mean
field theory for the interaction between the chains, while taking into
account the interaction inside a single chain exactly. Without the
interchain interaction the transition temperature had to be zero.
Introducing the interchain interaction and taking it into account in the
mean field approximation we have derived for the critical temperature a mean
field equation containing two-particles correlation functions for a single
isolated chain.

Although this was a considerable simplification, methods of calculation of
two-particles correlation functions in 1D had not been developed for an
arbitrary interaction (Tomonaga-Luttinger models with long-range interaction
was an exception). Well-developed Bethe Ansatz methods did not allow to
calculate the correlation functions and the problem looked quite
non-trivial, although solutions had been found in models with a linear
spectrum and with a long-range interaction \cite{dzlar,luthpesch}, as well
as with a special value of the backward scattering \cite{luthem}.

In Ref. \cite{eflar1974} we have calculated correlation functions of
superconducting order parameters in wires considering gapless fluctuations
of the phase $\phi ,$ while in Ref. \cite{efetov1975} calculated correlation
functions of a one-dimensional Fermi gas with a strong attraction. The srong
attraction lead to formation of bosonic electron pairs with a repulsion
between them. Using a Jordan-Wigner transformation we reduced the
thermodynamics of the system to the one of spinless fermions, while the
correlation function of the superconducting order parameters was written in
terms of a Toeplitz determinant. We have calculated these functions at large
distances or times and compared them with those calculated in different
models in Refs. \cite{dzlar,eflar1974,luthpesch,luthem}. As a result, we
discovered that all of them had the form
\begin{equation}
\Pi _{s}\left( R,\tau \right) \varpropto \frac{T^{\alpha }}{\left\vert \sinh
\pi T\left( R/v_{s}+i\tau \right) \right\vert ^{\alpha }},  \label{a1}
\end{equation}%
where $R$ is the distance between two points in space, $\tau $ is the
distance in the imaginary time, $T$ is temperature, $v_{s}$ is the velocity
of excitations, and $\alpha $ is a constant depending on the model
considered. All this has allowed us to propose a hypothesis that the form of
the correlation functions at large distances and times is formed by gapless
excitations, and in order to calculate, e.g., the superconducting
correlation function
\begin{equation}
\Pi _{s}\left( R,\tau \right) =\left\langle \psi _{\uparrow }^{+}\left(
R,\tau \right) \psi _{\downarrow }^{+}\left( R,\tau \right) \psi
_{\downarrow }\left( 0,0\right) \psi _{\uparrow }\left( 0,0\right)
\right\rangle ,  \label{a2}
\end{equation}%
where $\psi ^{+}\left( x,\tau \right) $ and $\psi \left( x,\tau \right) $
are creation and anihillation operators, one has to replace the pairs of the
operators by the following operator
\begin{equation}
\psi _{\uparrow }^{+}\left( x,\tau \right) \psi _{\downarrow }^{+}\left(
x,\tau \right) \rightarrow ae^{2i\hat{\phi}\left( R,\tau \right) },
\label{a3}
\end{equation}%
where $a$ is a constant, and calculate the correlation function $\Pi
_{s}\left( R,\tau \right) $ representing it in the form%
\begin{equation}
\Pi _{s}\left( R,\tau \right) \propto \left\langle e^{2i\hat{\phi}\left(
R,\tau \right) }e^{-2i\hat{\phi}\left( 0,0\right) }\right\rangle _{eff}.
\label{a4}
\end{equation}%
The angular brackets in Eq. (\ref{a4}) stand for quantum mechanical
averaging with an effective Hamiltonian
\begin{equation}
\hat{H}_{eff}=\frac{1}{2}\int \left[ \frac{\left( \hat{\rho}\left( x\right) -%
\bar{\rho}\right) ^{2}}{K}+Kv_{s}^{2}\left( \frac{\partial \hat{\phi}\left(
x\right) }{\partial x}\right) ^{2}\right] dx.  \label{a5}
\end{equation}%
In Eq. (\ref{a5}) $\hat{\phi}\left( x\right) $ and $\hat{\rho}\left(
x\right) $ are the phase and density operators satisfying the following
commutations relations
\begin{equation}
\left[ \hat{\rho}\left( x\right) ,\hat{\phi}\left( x^{\prime }\right) \right]
=\delta \left( x-x^{\prime }\right) ,  \label{a6}
\end{equation}%
the constant $K=\partial \bar{\rho}/\partial \mu $ is compressibility ($\bar{%
\rho}$ -is the electron density). The imaginary time dependence of the
operators $\hat{\phi}\left( x,\tau \right) $ and $\hat{\rho}\left( x,\tau
\right) $ is determined by usual quantum mechanical relations%
\begin{eqnarray}
\hat{\phi}\left( x,\tau \right) &=&e^{H_{eff}\tau }\hat{\phi}\left( x\right)
e^{-H_{eff}\tau },\quad  \label{a7} \\
\hat{\rho}\left( x,\tau \right) &=&e^{H_{eff}\tau }\rho \left( x\right)
e^{-H_{eff}\tau }.  \nonumber
\end{eqnarray}%
Calculating the correlation function $\Pi _{s}\left( R,\tau \right) $, Eq. (%
\ref{a2}), with the Hamiltonian $\hat{H}_{eff}$, Eq. (\ref{a5}), one comes
to Eq. (\ref{a1}) with
\begin{equation}
\alpha =2\left( \pi Kv_{s}\right) ^{-1}.  \label{a8}
\end{equation}%
Remarkably, one can calculate in the same way not only the superconducting
correlations but also correlation function $\Pi _{d}\left( R,\tau \right) $
of $2p_{F}$-components $\tilde{\rho}\left( x,\tau \right) $ of the electron
density. One introduces this function as
\begin{equation}
\Pi _{d}\left( R,\tau \right) \propto \sum_{\alpha ,\beta }\psi _{\alpha
}^{+}\left( R,\tau \right) \psi _{\alpha }\left( R,\tau \right) \psi _{\beta
}^{+}\left( 0,0\right) \psi _{\beta }\left( 0,0\right) -\bar{\rho}^{2}
\label{a9}
\end{equation}

Then, one can make a replacement%
\begin{equation}
\psi _{\alpha }^{+}\left( x,\tau \right) \psi _{\alpha }\left( x,\tau
\right) \rightarrow b\cos \left( \pi \int \hat{\rho}\left( x\right)
dx\right) .  \label{a10}
\end{equation}%
Calculation of the correlation function $\Pi _{d}\left( R,\tau \right) $,
Eqs. (\ref{a9}, \ref{a10}), is similar to the calculation of the correlation
function $\Pi _{s}\left( R,\tau \right) $, Eq. (\ref{a2}) and one comes to
the following formula
\begin{equation}
\Pi _{d}\left( R,\tau \right) \varpropto \frac{T^{1/\alpha }}{\left\vert
\sinh \pi T\left( R/v_{s}+i\tau \right) \right\vert ^{1/\alpha }},
\label{a15}
\end{equation}%
demonstrating an interesting duality with Eq. (\ref{a1}).

We have proposed that the constants $v_{s}$, $K$ and, hence, the exponent $%
\alpha $ can be calculated using the Bethe-Ansatz method. The knowledge of
the correlation functions $\Pi _{s}\left( R,\tau \right) $ and $\Pi
_{d}\left( R,\tau \right) $ has allowed us to estimate the transition
temperatures to both superconducting and dielectric states in
quasi-one-dimensional systems. Our hypothesis about the crucial role of the
gapless excitations has in fact been confirmed later by the development of
the theory of the Luttinger liquid by Haldane \cite{haldane}. That theory is
also based of considering gapless excitations in 1D systems but is more
rigorous and allows one to calculate many details.

The superconducting transition temperature $T_{c}$ can be calculated in the
mean field approximation with respect to different chains. The mean field
equation can be written within this scheme as
\begin{equation}
1=W\int \int_{0}^{1/T}\Pi _{s}\left( R,\tau \right) dRd\tau ,  \label{a18}
\end{equation}%
where $W$ is proportional to the square of the tunneling amplitude from
chain to chain. An equation for the transition into a charge density wave
can be written analogously using the correlation function $\Pi _{d}\left(
R,\tau \right) .$ One should keep in mind, though, that not only the
interchain tunnelling but also Coulomb interaction enters the mean field
equation.

\section{\label{sec:granular}Granular superconductors and Josephson neworks.}

The importance of the phase fluctuations of the superconducting order
parameter has inspired me to apply several years later the same idea for
description of granular superconductors. Certain granular materials could be
produced in that time, although they were not very homogeneous with respect
to the size of the grains and their arrangement. Artificially designed
neworks of Josephson junctions appeared considerably later but, from the
point of view of the theory developed in Ref. \cite{efetov1980}, these
systems are equivalent. At first glance, one could use the a similar scheme
as described in the previous section by replacing the quasi-one-dimensional
system by a `quasi-zero-dimensional' one.

However, in one-dimensional systems the long-range Coulomb interaction does
not play an important role. Everything is completely different in a granular
system. Indeed, in order to carry supercurrent a Cooper pair has to tunnel
from a grain to another grain. However, this change of the charge
configuration costs a considerable electrostatic energy and the supercurrent
can simply be blocked. Fortunately, the electrostatic Coulomb interaction
can rather easily be incorporated into an effective Hamiltonian containing
only phase fluctuations. In contrast to the one-dimensional systems where
theory based on the phase Hamiltonian $H_{eff}$, Eq. (\ref{a5}), has been
introduced semi-phenomenologically, the derivation of the proper Hamiltonian
for the granular system could be well justified unless the size of the
grains was too small, such that the condition%
\begin{equation}
\delta \ll \Delta _{0},  \label{a16}
\end{equation}%
where $\delta $ is the mean level spacing and $\Delta _{0}$ is the modulus
of the superconducting order parameter, could not be fulfilled.

Actually, each grain can be considered as zero dimensional with respect to
variations of the order parameter $\Delta .$ In other words, $\Delta =\Delta
_{0}e^{i\phi }$ does not depend on coordinates inside the grains but its
values vary from grain to grain. The Coulomb interaction can be taken into
account writing the electrostatic energy of the tunnelling of the Cooper
pair from grain $i$ to grain $j$ as
\begin{equation}
B_{ij}=2e\left( C^{-1}\right) _{ij},  \label{a17}
\end{equation}%
where $e$ is the electron charge, and $C_{ij}$ is the capacitance matrix.

Microscopic derivation leads to the following effective Hamiltonian $\hat{H}%
_{eff}$ containing only the phases $\phi _{i}$ of the order parameter in the
grain $i$%
\begin{equation}
\hat{H}_{eff}=\sum_{i,j}\left[ \frac{1}{2}B_{ij}\hat{\rho}_{i}\hat{\rho}%
_{j}+J_{ij}\left( 1-\cos \left( \phi _{i}-\phi _{j}\right) \right) \right] ,
\label{a20}
\end{equation}
where the operator $\hat{\rho}_{i}$ of the number of the Cooper pairs in the
grain $i$ equals
\begin{equation}
\hat{\rho}_{i}=-i\frac{\partial }{\partial \phi _{i}},  \label{a21}
\end{equation}%
and $J_{ij}$ are Josephson couplings between grains $i$ and $j.$ The
eigenvalues of the operators $\hat{\rho}_{i}$ are integers.

Again, the Hamiltonian $\hat{H}_{eff}$ describes low energy fluctuations of
the order parameter. Higher energy degrees of freedom has been integrated
out. The model described by the Hamiltonian $\hat{H}_{eff}$ is a version of
the quantum $XY$-model. It has been demonstrated \cite{efetov1980} making a
mean field theory for the model specified in Eq. (\ref{a20}) that at low
temperatures the system undergoes a superconductor-insulator transition. The
mean field equation has been written in the form%
\begin{equation}
1=\frac{J}{2}\int_{0}^{1/T}\Pi \left( \tau \right) d\tau ,  \label{a22}
\end{equation}%
where $\Pi \left( \tau \right) $ is the correlation function of the
superconducting order parameters%
\begin{equation}
\Pi \left( \tau \right) =\left\langle e^{i\phi \left( 0\right) }e^{-i\phi
\left( \tau \right) }\right\rangle _{0},  \label{a23}
\end{equation}%
$J=\sum_{j}J_{ij},\,$and
\[
\phi \left( \tau \right) =e^{i\hat{H}_{0}\tau }\phi e^{-i\hat{H}_{0}\tau
}.\quad
\]%
In Eqs. (\ref{a23}),
\begin{equation}
\hat{H}_{0}=\frac{1}{2}\sum_{i,j}B_{ij}\hat{\rho}_{i}\hat{\rho}_{j}
\label{a24}
\end{equation}%
is the effective Hamiltonian of isolated grains and the angular brackets $%
\left\langle ...\right\rangle _{0}$ stand for the averaging with this
Hamiltonian. Eqs. (\ref{a22}-\ref{a24}) show that the phase diagram is
completely determined by the correlation function of the order parameters $%
\Pi \left( \tau \right) .$ It is important that $\Pi \left( \tau \right) $
satisfies the bosonic periodic boundary condition
\begin{equation}
\Pi \left( \tau \right) =\Pi \left( \tau +1/T\right) .  \label{a25}
\end{equation}

Calculation of the average in Eq. (\ref{a23}) is not difficult and one comes
to the following formula%
\begin{eqnarray}
\Pi \left( \tau \right) &=&Z^{-1}\sum_{n_{1},n_{2},n_{3}..n_{N}}\exp \left[ -%
\frac{\tau B_{11}}{2}-\tau \sum_{j}B_{1j}n_{j}\right]  \nonumber \\
&&\times \exp \left[ -\frac{1}{2T}\sum_{i,j}B_{ij}n_{i}n_{j}\right] ,
\label{a26}
\end{eqnarray}%
where
\begin{equation}
Z=\sum_{n_{1},n_{2},n_{3}..n_{N}}\exp \left[ -\frac{1}{2T}%
\sum_{i,j}B_{ij}n_{i}n_{j}\right] .  \label{a27}
\end{equation}%
In Eqs. (\ref{a26}, \ref{a27}) summation is performed over $n_{i}=0,\pm
1,\pm 2,...$

One can see easily that the function $\Pi \left( \tau \right) $ satisfies
the periodicity condition (\ref{a25}). At $T=0,$ the correlation function $%
\Pi \left( \tau \right) $ reduces to a simple form
\begin{equation}
\Pi \left( \tau \right) =\exp \left[ -\frac{\tau B_{11}}{2}\right] .
\label{a28}
\end{equation}%
Substituting Eq. (\ref{a28}) into Eq. (\ref{a22}) one obtains the critical
value $J_{c}$ of the Josephson coupling at which the macroscopic
superconductivity appears%
\begin{equation}
J_{c}=E_{c}=\frac{B_{11}}{2}.  \label{a29}
\end{equation}%
In Eq. (\ref{a29}) the energy $E_{c}$ is the energy of adding a Cooper pair
into the grain. Nowadays, the effect of the destruction of superconductivity
or conductivity by the Coulomb interaction is known under the name `Coulomb
blockade'.

An interesting property of the function $\Pi \left( \tau \right) $ is that
it is periodic not only in the imaginary time, Eq. (\ref{a25}), but also in
the real one $t$. Making a Wick rotation $\tau \rightarrow it$ one obtains
for the function $K\left( t\right) =\Pi \left( it\right) ,$
\begin{eqnarray}
K\left( t\right) &=&Z^{-1}\sum_{n_{1},n_{2},n_{3}..n_{N}}\exp \left[ -\frac{%
itB_{11}}{2}-it\sum_{j}B_{1j}n_{j}\right]  \nonumber \\
&&\times \exp \left[ -\frac{1}{2T}\sum_{i,j}B_{ij}n_{i}n_{j}\right] .
\label{a30}
\end{eqnarray}%
In the limit of low temperatures $T\rightarrow 0$ one comes to a simple
formula
\begin{equation}
K_{0}\left( t\right) =\exp \left[ -\frac{itB_{11}}{2}\right]  \label{a31}
\end{equation}%
demonstrating the periodicity of the function $K_{0}\left( t\right) $ with
the period $t_{0}$,%
\begin{equation}
t_{0}=\frac{2\pi }{E_{c}}.  \label{a32}
\end{equation}%
In this limit one can consider the grain as a two-level systems (`no Cooper
pairs' and `one Cooper pair').

The real time correlation function determines a frequency dependent response
$K\left( \omega \right) $ to an external electric field. At $T=0,$ its
imaginary part has a $\delta $-functional form
\begin{equation}
ImK\left( \omega \right) \propto \delta \left( \omega -\frac{1}{2}\left(
B_{11}+B_{22}-2B_{12}\right) \right) .  \label{a33}
\end{equation}%
Eq. (\ref{a33}) can be used when the tunnelling between the two grains is
small but finite.

It is interesting to note that the same correlation function $\Pi \left(
\tau \right) $, Eq. (\ref{a26}), arises in grains fabricated from normal
metal \cite{eftscher}, which is not accidental because this is also an
effect of the charge quantization. One can read more about the granular
electron systems in the review \cite{beloborodov}.

\section{\label{sec:supersym}Supersymmetry in disorder and chaos.}

\subsection{Prehistory.}

The prediction of a new phenomenon of the Anderson localization \cite%
{anderson} has strongly stimulated both theoretical and experimental study
of disordered materials. At the same time, one could see from that the
Anderson's work that quantitative description of the disordered systems was
not a simple task and many conclusions were based on semi-qualitative
arguments. Development of theoretical methods for quantitative study of
quantum effects in disordered systems was clearly very demanding.

The most straightforward way to take into account disorder is using
perturbation theory in the strength of the disorder potential \cite{agd}.
However, the phenomenon of the localization is not easily seen within this
method and the conventional classical Drude formula for conductivity was
considered in \cite{agd} as the final result for the dimensionality $d>1$.
This result is obtained after summation of diagrams without intersection of
impurity lines. Diagrams with intersection of the impurity lines give a
small contribution if the disorder potential is not strong, so that $%
\varepsilon _{0}\tau \gg 1,$ where $\varepsilon _{0}$ is the energy of the
particles (Fermi energy in metals) and $\tau $ in the elastic scattering
time.

Although there was a clear understanding that the diagrams with the
intersection of the impurity lines were not small for one dimensional
chains, $d=1$, performing explicit calculations for those systems was
difficult. This step has been done considerably later by Berezinsky \cite%
{berezinsky} who demonstrated localization of all states in 1D chains by
summing complicated series of the perturbation theory. As concerns higher
dimensional systems, $d>1$, the Anderson transition was expected at a strong
disorder but it was clear that the perturbation theory could not be applied
in that case.

So, the classical Drude theory was considered as a justified way of the
description of disordered metals in $d>1$ and $\varepsilon _{0}\tau \gg 1$.
At the same time, several results for disordered systems could not be
understood within this simple generally accepted picture.

In 1965 Gorkov and Eliashberg \cite{ge} suggested a description of level
statistics in small disordered metal particles using the random matrix
theory (RMT) of Wigner and Dyson \cite{wigner,dyson}. At first glance, the
diagrammatic method of Ref. \cite{agd} had to work for such a system but one
could not see any indication on how the formulae of RMT could be obtained
diagrammatically. Of course, the description of Ref. \cite{ge} was merely a
hypothesis and the RMT had not been used in the condensed matter before but
nowadays it looks rather strange that this problem did not attract an
attention. Apparently, the diagrammatic methods were not very widely used in
that time and therefore not so many people were interested in resolving such
problems.

Actually, the discrepancies were not discussed in the literature until 1979,
the year when the publication \cite{aalr} appeared. In this work,
localization of all states for any disorder already in $2D$ was predicted.
This result has attracted much attention and it was simply unavoidable that
people started thinking about how to confirm it diagrammatically. The only
possibility could be that there were some diverging quantum corrections to
the classical conductivity, and soon the mechanism of such divergencies has
been discovered \cite{glk,aar,ar}.

It turns out that the sum of a certain class of the diagrams with
intersecting impurity lines diverges in the limit of small frequencies $%
\omega \rightarrow 0$ in a low dimension $d\leq 2$ and it can be considered
as new effective mode. This mode has a form of the diffusion propagator and
its contribution to the conductivity $\sigma \left( \omega \right) $ can be
written in the form%
\begin{equation}
\sigma \left( \omega \right) =\sigma _{0}\left( 1-\frac{1}{\pi \nu }\int
\frac{1}{D_{0}\mathbf{k}^{2}-i\omega }\frac{d^{d}\mathbf{k}}{\left( 2\pi
\right) ^{d}}\right)  \label{b1}
\end{equation}%
where $D_{0}=v_{0}^{2}\tau /3$ is the classical diffusion coefficient and $%
\sigma _{0}=2e^{2}\nu D_{0}$ is the classical conductivity. The parameters $%
v_{0}$ and $\nu $ are the Fermi velocity and density of states on the Fermi
surface.

Similar contributions arise also in other quantities. Eq. (\ref{b1})
demonstrates that in the dimensions $d=0,1,2$ the correction to conductivity
diverges in the limit $\omega \rightarrow 0$. It is very important that the
dimension is determined by the geometry of the sample. In this sense, small
disordered particles correspond to zero dimensionality, $d=0,$ and wires to $%
d=1$.

In this way, one can reconcile the hypothesis about the Wigner-Dyson level
statistics in disordered metal particles and assertion about the
localization in thick wires and $2D$ films with the perturbation theory in
the disorder potential. The divergences due to the contribution of the
diffusion modes make the perturbation theory inapplicable in the limit $%
\omega \rightarrow 0$ and therefore one does not obtain just the classical
conductivity using this approach.

Unfortunately, the divergence of the quantum corrections to the conductivity
in the limit $\omega \rightarrow 0$ makes the direct analytical
consideration very difficult for small $\omega $ because even the summation
of all orders of the perturbation theory does not necessarily lead to the
correct result. For example, the formulae for the level-level correlation
functions \cite{wigner, dyson} contain oscillating parts that cannot be
obtained in any order of the perturbation theory.

All this meant that a better tool had to be invented for studying the
localization phenomena and quantum level statistics. Analyzing the
perturbation theory one could guess that a low energy theory explicitly
describing the diffusion modes rather than single electrons might be an
adequate method.

The first formulation of such a theory was proposed by Wegner \cite{wegner}
who has introduced a non-linear $\sigma $-model based on a replica
representation of electron Green functions in a form of functional integrals
over complex fields.

Working with this model one has to integrate over $N\times N$ matrices $Q$
obeying the constraint $Q^{2}=1$. The $\sigma $-model is renormalizable and
renormalization group equations were written in Ref. \cite{wegner}. These
equations agreed with the perturbation theory of Eq. (\ref{b1}) and with the
scaling hypothesis of Ref. \cite{aalr}.

However, the saddle point approximation was not carefully worked out in \cite%
{wegner} because the saddle points were in the complex plane, while the
original integration had to be done over the real axis. This question was
addressed in the subsequent publications \cite{sw,elk}.

In the work \cite{sw}, the initial derivation of Ref. \cite{wegner} was done
more carefully and the authors have come to the conclusion that the matrices
$Q$ varied on a hyperboloid.

In contrast, the derivation of Ref. \cite{elk} was based on a representation
of the electron Green functions in a form of an integral over Grassmann
anticommuting variables. As a result, the $\sigma $-model derived in Ref.
\cite{elk} has lead to the result that the $N\times N$ matrix $Q$ varied on
a sphere. As in all these works the replica trick was used, there was no
contradiction between the approaches because one expected that in the limit $%
N\rightarrow 0$ both the models would give the same results. Indeed, the
perturbation theory and renormalization group calculations lead to identical
formulas.

The compact replica $\sigma $-model of Ref. \cite{elk} has several years
later been extended by Finkelstein \cite{finkelstein} to models of
interacting electrons. An additional topological term was added to this
model by Pruisken \cite{pruisken} for studying the Integer Quantum Hall
Effect. So, after all, the compact replica $\sigma $-models have helped to
solve interesting problems in the localization theory.

I was excited by all this development in particular because the description
of the disorder problems could be reformulated in terms of field theories.
One could speak of the matrix $Q$ in terms of an order parameter and could
consider the diffusion modes as Goldstone modes arising due to degeneracy of
the ground state. This correlated very well with my tastes, and I decided to
move into this field instead of continuing study of granular superconductors
and Josephson networks.

\subsection{Supermatrix `order parameter'.}

However, everything turned out to be considerably more complicated for
non-perturbative calculations. Attempts to study the level-level statistics
in a limited volume and localization in disordered wires using the replica $%
\sigma $-model of Ref. \cite{elk} have lead me to the conclusion that the
replica $\sigma $-models were not a convenient tool for studying
non-perturbative problems.

At this point I would like to mention again the atmosphere at the Landau
Institute created by Isaak Markovich. I had spent a year trying to calculate
a level-level correlation function in a disordered metallic grain but I did
not feel any pressure to publish something and could continue my work.
Finally, it resulted in constructing another type of the $\sigma $-model
that was not based on the replica trick. I called the proposed technique
`supersymmetry method', although the word `supersymmetry' is often used in
field theory in a narrower sense. The field theory derived for the
disordered systems using this approach has the same form of the $\sigma $%
-model as the one obtained with the replica trick, and all perturbative
calculations are similar \cite{efetov1}.

The free energy functional $F\left[ Q\right] $ of the $\sigma $-model has a
standard form
\begin{equation}
F\left[ Q\right] =\frac{\pi \nu }{8}\int STr\left[ D_{0}\left( \nabla
Q\right) ^{2}+2i\left( \omega +i\delta \right) \Lambda Q\right] d\mathbf{r}
\label{b2}
\end{equation}%
where $D_{0}=v_{0}^{2}\tau /d$ is the classical diffusion coefficient ($%
v_{0} $ is the Fermi velocity and $d$ is the dimensionality of the sample)
and the $8\times 8$ supermatrix $Q$ obeys the constraint
\begin{equation}
Q^{2}=1,  \label{b3}
\end{equation}%
and $\omega $ is the frequency and $\nu $ is the average density of states.
The symbol $`Str^{\prime }$ stands for supertrace.

The matrix $\Lambda $ equals
\begin{equation}
\Lambda =\left(
\begin{array}{cc}
1 & 0 \\
0 & -1%
\end{array}%
\right) ,  \label{b4}
\end{equation}%
and the supermatrix $Q$ can be written in the form%
\[
Q=U\Lambda \bar{U},
\]%
where $\bar{U}U=1.$

The matrix $U$ has both compact sector on the sphere and non-compact one on
the hyperboloid. They are glued with each other by anticommuting Grassmann
fields.

Calculation of, e.g., density-density correlation function $K_{\omega
}\left( \mathbf{r}\right) $ reduces to calculation of a functional integral
over $Q$%
\begin{equation}
K_{\omega }\left( \mathbf{r}\right) =2\int Q_{\alpha \beta }^{12}\left(
0\right) Q_{\beta \alpha }^{21}\left( \mathbf{r}\right) \exp \left( -F\left[
Q\right] \right) DQ,  \label{b5}
\end{equation}%
while
\begin{equation}
\left\langle Q\right\rangle =\int Q\exp \left( -F\left[ Q\right] \right) DQ=1
\label{b6}
\end{equation}%
is proportional to the average density of states.

The level-level correlation function $R\left( \omega \right) $ is given by
relation%
\begin{equation}
R\left( \omega \right) =\frac{1}{2}-\frac{1}{2}Re\int
Q_{11}^{11}Q_{11}^{22}\exp \left( -F_{0}\left[ Q\right] \right) dQ,
\label{b7}
\end{equation}%
where $\omega $ is the distance between the levels. One should also notice
absence of a weight denominator in Eqs. (\ref{b5}, \ref{b6}) that simply
equals to unity due to the supersymmetry. The superscripts of the matrix $Q$
stand for blocks that are in the same space as those in Eq. (\ref{b4}) and
the subscripts numerate elements in these blocks.

Eqs. (\ref{b5}, \ref{b6}, \ref{b7}) display a reformulation of the initial
problem of disordered metals in terms of a field theory that does not
contain disorder because the averaging over the initial disorder has already
been carried out. The latter enters the theory through the classical
diffusion coefficient $D_{0}$. The supermatrix $\sigma $-model described by
Eq. (\ref{b5}) resembles $\sigma $-models used for calculating contributions
of spin waves for magnetic materials. At the same time, the non-compactness
of the symmetry group of the supermatrices $Q$ makes this $\sigma $-model
unique.

We see that the supermatrix $Q$ plays to some extent the role of an order
parameter and its fluctuations are similar to Goldstone modes. At the same
time, its average is a constant and, actually, the Anderson metal-insulator
transition occurs as a result of fluctuations.

The first attempt to calculate the level-level correlation function lead to
a real surprise: the method worked \cite{efetov2}. For example, considering
the unitary ensemble one could reduce calculation of the integral in Eq. (%
\ref{b7}) to the following integral over two variables
\begin{equation}
R\left( \omega \right) =1+\frac{1}{2}Re\int_{1}^{\infty }\int_{-1}^{1}\exp %
\left[ i\left( x+i\delta \right) \left( \lambda _{1}-\lambda \right) \right]
d\lambda _{1}d\lambda  \label{b9}
\end{equation}%
where $x=\pi \omega /\Delta $, and $\Delta $-is the mean level spacing.
Calculations for the orthogonal and symplectic ensembles could be reduced to
integrals over three variables. The remaining integration is trivial for the
unitary ensemble, Eq. (\ref{b9}), and doable for the orthogonal and
symplectic ensembles leading to the famous formulae for level-level
correlation functions
\begin{equation}
R_{orth}\left( \omega \right) =1-\frac{\sin ^{2}x}{x^{2}}-\frac{d}{dx}\left(
\frac{\sin x}{x}\right) \int_{1}^{\infty }\frac{\sin xt}{t}dt  \label{b10}
\end{equation}%
\begin{equation}
R_{unit}\left( \omega \right) =1-\frac{\sin ^{2}x}{x^{2}}  \label{b11}
\end{equation}%
\begin{equation}
R_{sympl}\left( \omega \right) =1-\frac{\sin ^{2}x}{x^{2}}+\frac{d}{dx}%
\left( \frac{\sin x}{x}\right) \int_{0}^{1}\frac{\sin xt}{t}dt  \label{b12}
\end{equation}%
Eqs. (\ref{b10}- \ref{b12}) are known in the Wigner-Dyson theory \cite%
{wigner,dyson}, this result established the relevance of the latter to the
disordered systems. Since then one could use the RMT for calculations of
various physical quantities in mesoscopic systems or calculate directly
using the zero-dimensional supermatrix $\sigma $-model. Actually, to the
best of my knowledge, this was the first explicit demonstration that RMT
could correspond to a real physical system. Its original application to
nuclear physics was in that time phenomenological and confirmed by neither
analytical nor numerical calculations.

A direct derivation of Eqs. (\ref{b10}-\ref{b12}) from gaussian ensembles of
the random matrices using the supermatrix approach was done in the review
\cite{vwz}. This allowed the authors to compute certain average
compound-nucleus cross sections that could not be calculated using the
standard RMT route.

The proof of the applicability of the RMT to the disordered systems was
followed by the conjecture of Bohigas, Giannonni and Schmid \cite{bgs} about
the possibility of describing by RMT the level statistics in classically
chaotic clean billiards. Combination of the results for clean and disordered
small systems (billiards) has established the validity of the use of RMT in
mesoscopic systems. Some researches use for explicit calculations methods of
RMT but many others use the supermatrix zero-dimensional $\sigma $-model
(for review see, e.g. \cite{gmw,suptrace,been}). At the same time, the $%
\sigma $-model is applicable to a broader class of systems than the
Wigner-Dyson RMT because it can be used in higher dimensions as well.
Actually, one can easily go beyond the zero dimensionality taking higher
space harmonics in $F\left[ Q\right] $, Eq. (\ref{b2}). In this case, the
universality of Eqs. (\ref{b10}-\ref{b11}) is violated. One can study this
limit for $\omega \gg \Delta $ using also the standard diagrammatic
expansions of Ref. \cite{agd} and this was done in Ref. \cite{altshk}.

The calculation of the level correlations in small disordered systems
followed by the full solution of the localization problem in wires \cite%
{efetov3}, on the Bethe lattice and in high dimensionality \cite%
{efetov84,zirnbauerbethe,efetov87,efetov87a,efetov88}. After that it has
become clear that the supersymmetry technique is really an efficient tool
suitable for solving various problems of theory of disordered metals.

By now several reviews and a book have been published \cite%
{efetov1983,vwz,book,gmw,mirlin,evers,anderson50} where numerous problems of
disordered, mesoscopic and ballistic chaotic system are considered and
solved using the supersymmetry method. The interested reader can find all
necessary references in those publications.

\section{\label{sec:time}Thermodynamic quantum time-space crystal.}

Space and time play in many respects a similar role in modern physics. At
the same time, many materials have stable crystalline structures that are
periodic in space but not in time. Are thermodynamic states with a periodic
time dependence of physical quantities forbidden by fundamental laws of
nature?

Several years ago Wilczek \cite{wilczek} proposed a concept of Quantum Time
Crystals using a rather simple model that possessed a state with a current
oscillating in time. Later a more careful consideration of the model \cite%
{bruno} has led to the conclusion that this was not an equilibrium state.
These publications were followed by a hot discussion of the possibility of
realization of a thermodynamically stable quantum time crystal \cite%
{wilczek1,li,bruno1,bruno2,nozieres,wilczek2}. More general arguments
against thermodynamically stable quantum time crystals have been presented
later \cite{watanabe}. As a result, a consensus has been achieved that
thermodynamically macroscopic quantum time crystals could not exist.

Slowly decaying oscillations in systems out of equilibrium were not
forbidden by the `no-go' theorems, and their study is definitely interesting
by its own. Recent theoretical \cite{sacha,sondhi1,sondhi2,nayak,yao} and
experimental \cite{autti,zhang,choi} works have clearly demonstrated that
this research field is very interesting and is fast growing. At present, the
term `Quantum Time Crystal' is usually used for non-equilibrium systems.

However, it turns out that thermodynamically stable quantum time crystals
are nevertheless possible. The results of \cite{bruno2,watanabe} are correct
for the models considered there but it was implied that the wave functions
of the ground state did not depend on time. Although this assumption is
usually correct in the thermodynamic equilibrium, time dependence of the
wave functions in the equilibrium is generally not forbidden.

Actually, a phase transition into a state with an order parameter
oscillating in both imaginary $\tau $ and real $t$ time is possible in a
model of interacting fermions, and this is demonstrated in this section.
Again, I start by introducing an order parameter $b$ but, in contrast to the
previous sections, I do not consider any fluctuations. All calculations are
performed in the mean field approximation but the order parameter of a
thermodynamically stable state oscillates in time. Here I give a short
account of ideas and results, while a more detailed presentation is given in
Ref. \cite{efetov2019}.

I start with a model with a Hamiltonian $\hat{H}$ already adopted for using
the mean field approximation

\begin{eqnarray}
\hat{H} &=&\sum_{p}c_{p}^{+}\left( \varepsilon _{p}^{+}+\varepsilon
_{p}^{-}\Sigma _{3}\right) c_{p}  \label{c1} \\
&&+\frac{1}{4V}\Big[\tilde{U}_{\mathrm{0}}\Big(\sum_{p}c_{p}^{+}\Sigma
_{1}c_{p}\Big)^{2}-U_{\mathrm{0}}\Big(\sum_{p}c_{p}^{+}\Sigma _{2}c_{p}\Big)%
^{2}\Big]  \nonumber
\end{eqnarray}

Equation (\ref{c1}) describes interacting fermions of two bands $1$ and $2$
, and $p=\left\{ \mathbf{p,}\alpha \right\} $ stands for the momentum $%
\mathbf{\ p}$ and spin $\alpha .$ The energies $\varepsilon _{a}\left(
\mathbf{p}\right) $ are two-dimensional spectra in the bands counted from
the chemical potential $\mu ,$ $\varepsilon _{p}^{\pm }=\frac{1}{2}\left(
\varepsilon _{1}\left( \mathbf{p}\right) \pm \varepsilon _{2}\left( \mathbf{p%
}\right) \right) ,$ the interaction constants $U_{\mathrm{0}}$ and $\tilde{U}%
_{\mathrm{0}}$ are positive, while $V$ is the volume of the system.
Two-component vectors $c_{p},$ $c_{p}^{+}$ contain creation and annihilation
operators for the fermions of the bands $1$ and $2$. Matrices $\Sigma
_{1},\Sigma _{2},\Sigma _{3}$ are Pauli matrices in the space of numbers $1$
and $2$ numerating the bands.

The Hamiltonian $\hat{H}$ written for the electron-hole pairs resembles the
Bardeen-Cooper-Schrieffer (BCS) Hamiltonian for Cooper pairs \cite{bcs}. At
the same time, one can imagine other physical systems described by Eq. (\ref%
{c1}). Hamiltonian $\hat{H}$ contains an inter-band attraction (term with
matrix $\Sigma _{2}$) and repulsion (term with $\Sigma _{1}$). Taking into
account only the term with the attraction one obtains in a spin-fermon model
introduced earlier \cite{volkov1,volkov2} static loop currents oscillating
in space with the double period of the lattice \cite{volkov3}. This
corresponds to a hypothetical d-density wave (DDW) state \cite{chakravarty}.

In order to obtain the new thermodynamic quantum time-space crystal state,
one should consider both the types of the interaction. It turns out \cite%
{efetov2019} in this case that the novel state with an order parameter
oscillating both in real and imaginary time is also possible in addition to
the DDW state. The order parameters of both the states oscillate in space
with a vector $\mathbf{Q=}\left( \pi ,\pi \right) $ corresponding to the
vector of antiferromagnetic modulations but their time dependence is
drastically different. In order to prove the existence of this state one
should calculate the free energy of the new state and compare it with that
of the DDW state.

Thermodynamics of quantum systems can be very conveniently studied using
imaginary time $\tau $ in the interval $\left( 0,1/T\right) $ \cite{agd} and
I sketch here the main steps of the calculation of the free energy. The
real-time behavior will be derived using a Wick rotation $\tau \rightarrow
it.$

Following the mean field theory scheme one introduces imaginary-time order
parameters $b\left( \tau \right) $ and $b_{1}\left( \tau \right) $
corresponding to the two interaction terms in Eq. (\ref{c1}), and computes
trace over fermionic states. Making a rotation
\begin{equation}
c_{p}\left( \tau \right) =\mathcal{U}_{0}\tilde{c}_{p}\left( \tau \right)
,\quad \mathcal{U}_{0}=\frac{1}{\sqrt{2}}\left(
\begin{array}{cc}
1 & i \\
i & 1%
\end{array}%
\right)   \label{c1a}
\end{equation}%
and using the relations
\begin{equation}
\mathcal{U}_{0}^{+}\Sigma _{2}\mathcal{U}_{0}=\Sigma _{3},\;\mathcal{U}%
_{0}^{+}\Sigma _{3}\mathcal{U}_{0}=-\Sigma _{2},\quad \mathcal{U}%
_{0}^{+}\Sigma _{1}\mathcal{U}_{0}=\Sigma _{1},  \label{1b}
\end{equation}%
one can write the free energy functional $\mathcal{F}$ of the model
described by the Hamiltonian $\hat{H}$ (\ref{c1}) in the form
\begin{eqnarray}
\frac{\mathcal{F}\left[ b,b_{1}\right] }{T} &=&\int_{0}^{1/T}\Big[-2\sum_{%
\mathbf{p}}\mathrm{tr}\left[ \ln \left( h\left( \tau ,\mathbf{p}\right)
-ib_{1}\left( \tau \right) \Sigma _{1}\right) \right] _{\tau ,\tau }
\nonumber \\
&&+V\left( \frac{b^{2}\left( \tau \right) }{U_{\mathrm{0}}}+\frac{%
b_{1}^{2}\left( \tau \right) }{\tilde{U}_{\mathrm{0}}}\right) \Big]d\tau ,
\label{c2}
\end{eqnarray}%
where
\[
h\left( \tau ,\mathbf{p}\right) =\partial _{\tau }+\varepsilon ^{+}\left(
\mathbf{p}\right) -\varepsilon ^{-}\left( \mathbf{p}\right) \Sigma
_{2}-b\left( \tau \right) \Sigma _{3},
\]%
symbol `$\mathrm{tr}$' means trace in space of the bands $1,2$, and
\begin{equation}
b\left( \tau \right) =b\left( \tau +1/T\right) ,\;b_{1}\left( \tau \right)
=b_{1}\left( \tau +1/T\right) .  \label{c3}
\end{equation}

The form of the interaction between the electron-hole pairs in Hamiltonian $%
\hat{H}$ (\ref{c1}) makes the mean field theory exact. Both the terms in the
functional $\mathcal{F}\left[ b,b_{1}\right] $ are proportional to the
volume $V$, and one can obtain the physical free energy simply minimizing $%
\mathcal{F}\left[ b,b_{1}\right] $ with respect to $b\left( \tau \right) $
and $b_{1}\left( \tau \right) .$ Although $b\left( \tau \right) =\gamma
,\;b_{1}\left( \tau \right) =0$ obtained previously \cite{volkov3} provides
a minimum of $\mathcal{F}\left[ b,b_{1}\right] $, there is a region of
parameters where the absolute minimum is reached at $\tau $-dependent
functions $b\left( \tau \right) $ and $b_{1}\left( \tau \right) .$
Unfortunately, the exact minimization is difficult and certain additional
approximations will be used.

Let us assume for a while that $b_{1}\left( \tau \right) =0$ and study
extrema of $\mathcal{F}\left[ b(\tau ),0\right] .$ Varying the functional $%
\mathcal{F}\left[ b\left( \tau \right) ,0\right] $ one comes to the
following equation

\begin{equation}
b\left( \tau \right) =-U_{0}\mathrm{tr}\int \Sigma _{3}\left[ h^{-1}\left(
\tau ,\mathbf{p}\right) \right] _{\tau ,\tau }\frac{d\mathbf{p}}{\left( 2\pi
\right) ^{2}}.  \label{c4}
\end{equation}%
Although Eq. (\ref{c4}) is quite non-trivial due to a possible dependence of
$b\left( \tau \right) $ on $\tau $, solutions $b_{0}\left( \tau \right) $
can be written exactly in terms of a Jacobi double periodic elliptic
function $\mathrm{sn}\left( x|k\right) $,
\begin{equation}
b_{0}\left( \tau \right) =k\gamma \mathrm{sn}\left( \gamma \left( \tau -\tau
_{0}\right) |k\right) ,  \label{c5}
\end{equation}%
where parameter $k,$ $0<k<1,$ is the modulus, $\gamma $ is an energy, and $%
\tau _{0}$ is an arbitrary shift of the imaginary time in the interval $%
0<\tau _{0}<1/T$ . In the limit $k\rightarrow 1$, the function has an
asymptotic behavior $\mathrm{sn}\left( x|k\right) \rightarrow \pm \tanh x,$
while in the limit $k\ll 1$ one obtains $\mathrm{sn}\left( x|k\right)
\rightarrow \sin x.$

The period of the oscillations for an arbitrary $k$ equals $4K\left(
k\right) /\gamma ,$ where $K\left( k\right) $ is the elliptic integral of
the first kind, and therefore the condition
\begin{equation}
\gamma =4K\left( k\right) mT  \label{c6}
\end{equation}%
with integer $m$ must be satisfied to fulfill equations (\ref{c3}). In the
most interesting limit of small $1-k,$ the period $4K\left( k\right) /\gamma
$ of $b_{0}\left( \tau \right) $ grows logarithmically as $\frac{1}{2}\ln $ $%
\left( \frac{8}{1-k}\right) ,$ and the solution $b_{0}\left( \tau \right) $
consists of $2m$ well separated alternating instantons and anti-instantons
with the shape $\pm \gamma \tanh \gamma \tau $. It is important that the
integral over the period of the oscillations in equation (\ref{c5}) equals
zero. Averaging over the position $\tau _{0}$ of the instanton one obtains
zero as well
\begin{equation}
\overline{b_{0}\left( \tau \right) }=0.  \label{c7}
\end{equation}

The existence of the non-trivial local minima $\mathcal{F}\left[ b,0\right] $%
, Eq. (\ref{c2}), at $b_{0}\left( \tau \right) $ has been established
previously by Mukhin \cite{mukhin,mukhin1,mukhin2} starting from a different
model. Generally, there can be many solutions corresponding to different
minima of $\mathcal{F}\left[ b,0\right] $ depending on the number $m$ of
instanton-antiinstanton pairs (IAP). However, the lowest value of the
functional $\mathcal{F}\left[ b,0\right] $ is reached at $m=0$ corresponding
to the static order \cite{efetov2019}. Coordinate-dependent Jacobi elliptic
functions are also solutions of a mean field time-independent equation
arising in 1D models of polymers, which has been discovered long ago \cite%
{brazovskii}.

The field $b_{1}\left( \tau \right) $ does not couple to the static order in
the mean field approximation, and this is why this field was not considered
previously \cite{volkov3}. The absence of the coupling is quite natural
because the order parameter $b\left( \tau \right) $ describes loop currents
oscillating in space with $\mathbf{Q=}\left( \pi ,\pi \right) ,$ while $%
b_{1}\left( \tau \right) $ corresponds to a charge oscillations with the
same vector. The situation changes when the field $b\left( \tau \right) $
varies in time. In order to understand what happens, one should calculate a
linear term in $b_{1}\left( \tau \right) $ in the expansion of $\mathcal{F}%
\left[ b,b_{1}\right] $, Eq. (\ref{c2}), in $b_{1}\left( \tau \right) $ for $%
b\left( \tau \right) =b_{0}\left( \tau \right) .$

Substituting $b_{0}\left( \tau \right) $ into Eq. (\ref{c2}) and expanding $%
\mathcal{F}\left[ b,b_{1}\right] $ in $b_{1}\left( \tau \right) $ one
obtains the following linear term
\begin{equation}
\frac{\mathcal{F}_{\mathrm{int}}\left[ b_{1}\right] }{VT}=-\frac{J}{2}%
\int_{0}^{1/T}\dot{b}_{0}\left( \tau \right) b_{1}\left( \tau \right) d\tau ,
\label{c8}
\end{equation}%
where $J$ is a constant.

Fluctuations of $b_{1}\left( \tau \right) $ generate an effective attraction
between the instantons and anti-instantons and favor formation of $\tau $
-dependent structures. The mechanism of the attraction is similar to the one
of the electron-phonon interaction in solids, which can be seen after a
formal replacement of $\tau $ by a coordinate. The field $b_{1}\left( \tau
\right) $ plays in this picture the role of phonons, and its fluctuations
may result in a sufficiently strong attraction of instantons and
anti-instantons. Eventually, an order parameter $b\left( \tau \right) $
depending on the imaginary time $\tau $ can provide the minima of the free
energy.

Exact minimization of the free energy functional $\mathcal{F}\left[ b,b_{1}%
\right] $ (\ref{c2}) is difficult. Therefore we simplify the study by
considering the limit of low temperatures $T$ when one can expect a large
number of IAP in the system and of small $1-k$ corresponding to a large
period of the IAP lattice. In this limit, the difference $\Delta F$ between
the total free energy $F$ and the free energy $F_{\mathrm{st}}$ of the
system with the static order parameter is proportional to $2m.$ The case $%
\Delta F/\left( 2mTV\right) >0$ corresponds to the state with the static
order parameter, while in the region of parameters where $\Delta F/V\left(
2mT\right) <0$ one expects a chain of alternating instantons and
anti-instantons.

Computation of the free energy is performed choosing
\begin{equation}
\varepsilon _{1}\left( \mathbf{p}\right) =\alpha p_{x}^{2}-\beta
p_{y}^{2}+P-\mu ,\quad \varepsilon _{2}\left( \mathbf{p}\right) =\alpha
p_{y}^{2}-\beta p_{x}^{2}-P-\mu  \label{c9}
\end{equation}%
corresponding to the spectrum of cuprates near the middle of the edges of
the Brillouin zone (momenta $\mathbf{p}$ are counted from the middle of the
edges), where $P$ is a Pomeranchuk order parameter obtained previously in a
spin-fermion model with overlapping hot spots \cite{volkov1}, and $\mu $ is
the chemical potential.

Numerical calculation of the free energy $\Delta F$ has been performed
expanding the free energy functional $\mathcal{F}\left[ b,b_{1}\right] $ up
to the second order in $b_{1}\left( \tau \right) $ and $\delta b\left( \tau
\right) =b\left( \tau \right) -b_{0}\left( \tau \right) ,$ and finding the
minimum of the quadratic form of these variables. It has been demonstrated
in Ref. \cite{efetov2019} that for small $1-k$ there is a region of
parameters where the free energy $\Delta F$ becomes negative, which
indicates that the time-independent DDW state is unstable. In this region a
chain of alternating instatons and antiinstantons appears. The number $m$ of
the instanton-antiinstanton pairs depends on temperature $T$ but this
dependence has has not been determined sofar.

Here, structures periodic in space (oscillations with vector $\mathbf{Q}%
_{AF} $ connecting the bands $1$ and $2$) are considered. Therefore the
periodic in $\tau $ order parameter $b\left( \tau \right) $ providing the
minimum of the free energy is at the same time the amplitude of the periodic
oscillations in space. As the present consideration does not determine the
number of the pairs as a function of temperature, we calculate physical
quantities without specifying the value of $m$ or $k$ related to each other
by Eq. (\ref{c6}).

The periodic structure described by the Jacobi elliptic function $%
b_{0}\left( \tau \right) $ (\ref{c5}) is actually double periodic in the
complex plane of $\tau $ and, hence, is periodic in real time $t.$
Remarkably, $b_{0}\left( it\right) $ still satisfies Eq. (\ref{c4}) after
the rotation $\tau \rightarrow it.$

It is convenient to introduce a function $B\left( t\right) ,$%
\begin{equation}
B\left( t\right) =b\left( it\right) .  \label{c10}
\end{equation}

The Jacobi elliptic function $\mathrm{sn}\left( iu,k\right) $ of an
imaginary argument $iu$ is related to an antisymmetric elliptic function $%
\mathrm{sc}\left( u|k\right) $ with the period $2K\left( k\right) $ as \cite%
{as}
\[
\mathrm{sn}\left( iu|k\right) =i\mathrm{sc}\left( u|k^{\prime }\right)
,\;k^{2}+k^{\prime 2}=1.
\]%
Therefore, the order parameter $B\left( t\right) $ is an imaginary and
antisymmetric in time (counted from $t_{0}$) function, while the function $%
B_{1}\left( t\right) =b_{1}\left( it\right) $ is real and symmetric. One can
write $B\left( t\right) $ in the form
\begin{equation}
B\left( t\right) =b\left( it\right) =i\gamma k\mathrm{sc}\left( \gamma
\left( t-t_{0}\right) |k^{\prime }\right) ,  \label{c11}
\end{equation}%
where $t_{0}$ is an arbitrary shift of time.

The oscillating behavior of the function $B\left( t\right) $ leads to
oscillations of wave functions. In order to calculate observable physical
quantities one should average the result over $t_{0}$ and this gives
immediately
\begin{equation}
\overline{B\left( t\right) }=0,  \label{c12}
\end{equation}%
which means that the average order parameter vanishes.

In order to calculate a $2$-times correlation function
\begin{equation}
N\left( t\right) =\overline{B\left( t\right) B^{\ast }\left( 0\right) }
\label{c13}
\end{equation}%
one can use a Fourier series for the function $\mathrm{sc}\left( u|k\right) $
. Then, the average over $t_{0}$ leads in the limit $1-k\ll 1$ to the
following result \cite{efetov2019}
\begin{equation}
N\left( t\right) \approx 2\gamma ^{2}\sum_{n=1}^{\infty }f_{n}^{2}\cos
\left( 2\gamma nt\right) ,\quad  \label{c14}
\end{equation}%
where%
\begin{equation}
f_{n}=\left[ 1-\frac{1}{2}\left( \frac{1-k}{8}\right) ^{2n}\right] .
\label{c15}
\end{equation}

Function $N\left( t\right) $ shows an oscillating behavior with the
frequencies $2\gamma n$ (we put everywhere $\hbar =1$). The energy $2\gamma $
is the energy of the breaking of electron-hole pairs and one can interpret
the form of $N\left( t\right) $ as oscillations between the static order and
normal state. The oscillations of $N\left( t_{1}-t_{2}\right) $ resemble
those of the order parameter in the non-equilibrium superconductors \cite%
{vk,spivak,barankov,altshuler,altshuler1,dzero,moor} but, in contrast to the
latter, the function $N\left( t\right) $ does not decay in time. The
contribution of high harmonics $n$ does not decay with $n$ but apparently
this is a consequence of the used approximations.

Non-decaying time oscillations of the two-time correlation function $N\left(
t\right) $ together with the vanishing of the single time average (\ref{c12}%
) allow us to generalize the definition of a space crystal by including time
in addition to the space coordinates. Therefore the physical state found
here can be classified as `Thermodynamic Quantum Time-Space Crystal'.

The correlation function $N\left( t\right) $, Eqs. (\ref{c14}, \ref{c15}),
was calculated by averaging over the position $t_{0}.$ Remarkably, the same
results for correlation functions can be obtained using an alternative
description based on the notion of an `operator order parameter' $\hat{B}.$
One can formally introduce a Hamiltonian $\hat{H}_{TC}$ \textrm{\ }for a
harmonic oscillator
\begin{equation}
\hat{H}_{TC}=2\gamma \left( a^{+}a+\frac{1}{2}\right) ,  \label{c16}
\end{equation}%
where $a^{+}$ and $a$ are bosonic creation and annihilation operators (for
simplicity, we consider here the limit $1-k\ll 1$). Using the Hamiltonian $%
\hat{H}_{TC}$ one can write the correlation function $N\left(
t_{1}-t_{2}\right) $ in the form
\begin{equation}
N\left( t\right) =\gamma ^{2}\left( \left\langle 0\left\vert A\left(
t\right) A^{+}\left( 0\right) \right\vert 0\right\rangle +\left\langle
0\left\vert A\left( 0\right) A^{+}\left( t\right) \right\vert 0\right\rangle
\right) ,  \label{c17}
\end{equation}

where
\[
A^{+}\left( t\right) =e^{i\hat{H}_{TC}t}A^{+}e^{-i\hat{H}_{TC}t},\;A^{+}=%
\sum_{n=1}^{\infty }f_{n}\frac{\left( a^{+}\right) ^{n}}{\sqrt{n!}}
\]%
and $\left\vert 0\right\rangle $ stands for the wave function of the ground
state of the Hamiltonian $\hat{H}_{TC}$ (\ref{c16}). At the same time,
quantum averages of the operators $A$ and $A^{+}$ vanish
\[
\left\langle 0\left\vert A\left( t\right) \right\vert 0\right\rangle
=\left\langle 0\left\vert A^{+}\left( t\right) \right\vert 0\right\rangle
=0.
\]%
One can interpret the operator $A$ as an operator order parameter. This type
of the order parameters extends the variety of conventional order parameters
like scalars, vectors, matrices used in theoretical physics. The
non-decaying time oscillations can be an important property for designing
qubits.

Possibility of an experimental observation depends on systems described by
the Hamiltonian (\ref{c1}). For cuprates, inelastic polarized neutron
spectroscopy can be a proper tool for observations. It is important that the
magnetic moments are basically perpendicular to the planes, which can help
to distinguish them from the antiferromagnetism spin excitations at $\left(
\pi ,\pi \right) .$ Calculating the Fourier transform $N\left( \omega
\right) $ of the function $N\left( t\right) $ and comparing it with the one
for the hypothetical time-independent DDW state $2\pi \gamma ^{2}\delta
\left( \omega \right) $ one can write at low temperatures the ratio of the
responses at $\left( \pi ,\pi \right) $ for these two states as
\begin{equation}
\chi \left( \omega ,\mathbf{q}\right) =\chi _{0}\sum_{n=1}^{\infty
}f_{n}\delta \left( \omega -2n\gamma \right) \delta \left( \mathbf{q-Q}%
_{AF}\right) ,  \label{c18}
\end{equation}%
where $\chi _{0}$ determines the response $\chi _{DDW}$ of the DDW state, $%
\chi _{DDW}\left( \omega \right) =\chi _{0}\delta \left( \omega \right) .$
In the absence of static magnetic moments the elastic scattering is not
expected to bring interesting information. Actually, anisotropic magnetic $%
\left( \pi ,\pi \right) $ excitations have been observed \cite{hayden} in $%
YBa_{2}Cu_{3}O_{6.9}$ but more detailed experiments are necessary to clarify
their origin.

The main conclusion of this section is the time-space crystals may exist as
a thermodynamically stable state in macroscopic systems. The order parameter
of the thermodynamic quantum crystals is periodic in both real and imaginary
times as well as in space but its average over the phases of the
oscillations vanish. The non-decaying oscillations can be seen, e.g., in
two-time correlation functions that determine cross-section in inelastic
scattering experiments. The frequency of the oscillations remains finite in
the limit of infinite volume, $V\rightarrow \infty $. One can expect various
experimental consequences and, in particular, one can suppose that the time
crystal may be a good candidate for the pseudogap state in superconducting
cuprates.

\section{\label{sec:conclusion}Conclusion.}

In this paper I tried to present rather different fields of research using a
generalized concept of the order parameter. In contrast to the standard
notion of a static long range order in an ordered phase, one may encounter
situations when there is no static long-range order. One can see from the
results of the investigation of several models of electrons with interaction
or moving in a random potential that there can be interesting non-trivial
physics. The properties of the models have been understood considering
either fluctuations or oscillations in space and time of a generalized order
parameter. Coulomb blockade, Anderson localization, space-time quantum
crystals, etc., are clearly quite different phenomena but their theoretical
description has many common features.

An essential part of my results presented in this paper either has been
done at the Landau Institute or followed from ideas developed there. I have
started my scientific carrier and worked for many years at the Landau
Institute at its best time, and I am personally very grateful to Isaac
Markovich for the creation of the Institute, for the support of my research,
and for giving me the possibility to work at the Institute.

Happy Birthday to you, Isaak Markovich! \bigskip

\textbf{Acknowledgements.} Financial support of Deutsche
Forschungsgemeinschaft (Projekt~EF~11/10-1) and of the Ministry of Science
and Higher Education of the Russian Federation in the framework of Increase
Competitiveness Program of NUST \textquotedblleft MISiS\textquotedblright
(Nr.~K2-2017-085") is greatly appreciated.

\end{document}